\documentstyle[12pt]{article}
\begin{document}
\newcommand{\ffa}{\stackrel{(1)}{\phi}}
\newcommand{\ffb}{\stackrel{(2)}{\phi}}
\newcommand{\ffc}{\stackrel{(3)}{\phi}}
\newcommand{\fa}{\stackrel{(1)}{\varphi}}
\newcommand{\fb}{\stackrel{(2)}{\varphi}}
\newcommand{\fc}{\stackrel{(3)}{\varphi}}
\newcommand{\n}{{\stackrel{(\alpha)}{n}}_i}
\renewcommand{\theequation}{\arabic{section}.\arabic{equation}}
\thispagestyle{empty}
\begin{flushright}
{\large \bf JINR preprint E2-93-181\\ Dubna,  1993} \end{flushright}
\vskip1.5cm

\begin{center}
{\large \bf On  squaring  the  primary  constraints  in  a   generalized
                     Hamiltonian dynamics}\\
\vskip0.8cm
{ \bf V.\ V.\ Nesterenko}\\[0.3cm]
{\it Laboratory of Theoretical Physics \\
Joint Institute for Nuclear Research,   Dubna \\
 SU-141980, Russia}\\[0.2cm]
E-mail address: nestr@theor.jinrc.dubna.su

\vskip2cm

                           {\bf Abstract}
\end{center}

     Consideration of  the  model  of the relativistic particle with
curvature and torsion in the three-dimensional space-time shows that
the squaring of the primary constraints entails a wrong result.  The
complete set of the Hamiltonian constraints arising here  correspond
to  another  model with an action similar but not identical with the
initial action.

\newpage
\vspace*{3cm}
\setcounter{section}{1}
     {\bf 1.}  The  generalized  Hamiltonian dynamics describing the
systems with constraints is widely used  now  in  investigating  the
theoretical  models  in  a contemporary elementary particle physics.
For example,  the gauge symmetries of various types,  without  which
every  model  does  not  practically  works,  inevitably  entail the
constraints in the phase space. Despite quite a large attention paid
to  the  Hamiltonian  systems  with  constraints (see,  for example,
papers [1--3] and  references  there  in)  some  topics  here  still
require  a  careful consideration.  The present note is dealing with
one of these problems,  namely,  the procedure of  squaring  of  the
primary  constraints  widely  used in practical calculations will be
investigated.  By making use of a concrete example,  a model of  the
socalled  relativistic  particle  with  curvature and torsion in the
three-dimensional  space-time  [4,  5],  we  will  show  that   this
procedure can result finally in an erroneous answer.

     The layout of the paper is the following. In the second section
high  lights  about  the  primary  constraints  are  given  and  the
procedure  of  squaring  of  these constraints is explained.  In the
third  section  a   generalized   Hamiltonian   description   of   a
relativistic  particle  with  curvature  and torsion is developed by
making use of the primary constraints in their original  form,  i.e.
in  the  form  that  follows  directly  from  the  definition of the
canonical momenta.  In the third section the Hamiltonian description
of this model is given by employing the squared primary constraints.
It is shown that in this case the  final  result  is  erroneous.  In
section  5  it is argued that the Hamiltonian formalism with the use
of  squared  primary  constraints  describes  in  the   case   under
consideration  another  model  with  an  action  analogous  but  not
identical with the initial action.

\setcounter{section}{2}
   {\bf 2.}   The  primary  constraints  are  a  starting  point  in
generating  a  complete  set  of  constraints   in   a   generalized
Hamiltonian  formalism  [1--3].  The  requirement  of preserving the
primary constraints  under  time  evolution  entails  the  secondary
constraints that in their turn should be preserved in time too. This
results in the tertiary constraints and so on.\footnote{According to
the  Dirac  terminology  [6]  all the constraints except the primary
ones are called the secondary constraints.}

     The  primary  constraints  follow   directly   from   the
definition of the canonical momenta
\begin{equation}
p_i(q,\,\dot q)\,=\,\frac{\partial L(q,\,\dot q)}{\partial \dot
q_i}\,{,}\quad i\,=\,1,\ldots , n.
\end{equation}
Given a degenerated  or a  singular Lagrangian  $L(q,\,\dot  q)$, the
functions $p_i(q,\,\dot q)$ obey $m\,=\,n-r$ constraints
\begin{equation}
\varphi _s(q,\,p)\,=\,0, \quad s\,=\,1, \dots ,m,
\end{equation}
where $n$ is the number of degrees of freedom  and  $r$  is  the
rank of the Hessian
\begin{equation}
\frac{\partial ^2L(q,\,\dot q)}{\partial \dot q_i\,\partial \dot q_j},
\quad 1\leq i,\,j\leq n.
\end{equation}
Upon substituting the functions $p_i(q,\,\dot q)$ in (2.2) by (2.1) all
the primary
constraints (2.2) vanish identically with respect to $q$ and $\dot q$.

In the case of the Lagrangian linear in velocities $r\,=\,0$ and
the definitions (2.1) are the primary constraints themselves
\begin{equation}
p_i\,=\,f_i(q), \quad i\,=\,1, \ldots ,n.
\end{equation}

Often it  turns out to be convenient to deal with primary constraints
preliminary
transformed instead of using them in their original form (2.1) or (2.4).
Squaring the left- and the right-hand side of (2.1)
and projecting this equation
onto suitable linearly independent vectors $\n (q),\;\;\alpha\,=\,
1,\,\ldots ,\,
m-1$ one obtains
$$
\sum_{i=1}^{n}p^2_i\,=\,\sum_{i=1}^{n}\left ( \frac{\partial L(q,\,\dot q)}
{\partial \dot q_i} \right )^2,
$$
\begin{equation}
\sum_{i=1}^n{}p_i\,\n(q)\,=\,\sum_{i=1}^{n}\frac{\partial L(q,\,\dot q)}
{\partial \dot q_i} \,\n (q),\quad \alpha \,=\,1,\,2,\, \ldots ,\,m-1.
\end{equation}
In the theory of the relativistic strings and membranes [7], for example,
this procedure enables one to get immediately relations
like (2.2) independent of the
velocities, i.e., the primary constraints in the Hamiltonian form.
However,  squaring
primary constraints does not prove to be always  correct, this will
be illustrated
further by  of a concrete example.

\setcounter{section}{3}
\setcounter{equation}0
Let us consider the model of the so-called relativistic particle
with curvature and torsion in the three-dimensional space-time. This model
is defined by the action [4, 5]
\begin{equation}
S\,=\,-\,m\int ds\,-\,\alpha\int k(s)\,ds \,-\,\beta \int \kappa (s)\,ds\,{,}
\end{equation}
where $\alpha $ and $\beta $ are dimensionaless parameters, $m$ is
a parameter
with the  dimension of mass, $ds$ is a differential of the length
of the world curve $x^\mu (s),\;\;\mu \,=\,0,\,1,\,2, \quad k(s)$ is the
curvature of this curve
\begin{equation}
k^2\,=\,-\,\frac{d^2x_\mu}{ds^2}\,\frac{d^2x^\mu }{ds^2}
\end{equation}
and $\kappa (s)$ is its torsion
\begin{equation}
\kappa(s)\,=\, k^{-2}\varepsilon _{\mu \nu \rho}x'^\mu x''^\nu x'''^\rho ,
\end{equation}
where $\varepsilon _{\mu \nu \rho}$ is a completely antisymmetric  unit
  tensor of the third rank,
$\varepsilon _{012}=+1$, the prime denotes the differentiation with
respect to $s$.
The Lorentz metric with signature $(+,\,-,\,-)$ is used.

The models of this kind have been
considered recently in  investigating the boson-fermion transformations
in external Chern-Simons fields [8--10], as the one dimensional version of
the rigid string [11, 12] and in
polymer physics [13].

Given an arbitrary parametrization of the world curve $x^\mu
(\tau),\;\;\mu \,=\,0,\,1,\,2$,
the action (3.1) can be rewritten as
$$
S\,=\,-m\int d\tau \sqrt{\dot x^2} -  \alpha \int d\tau
\frac{\sqrt{(\dot x\ddot x)^2 - \dot x^2\,\ddot x^2}}{\dot x^2} -
$$
\begin{equation}
-\,\beta \int d\tau \sqrt{\dot x^2}\,
\frac{\varepsilon_{\mu \nu \rho}\dot x^\mu \, \ddot x^\nu
\stackrel{\ldots}{x}^\rho}
{(\dot x\ddot x)^2\,-\,\dot x^2\,\ddot x^2},
\end{equation}
$$
 \dot x \equiv dx(\tau )/d\tau,\quad \dot x ^2>0,\quad D\,=\,3.
$$
It depends on the particle velocity, its acceleration, and on
the third derivatives of
the particle coordinates with respect to $\tau $. Therefore,
the canonical variables
are to be introduced according to Ostrogradsky [14, 15]
$$
q_1\,=\,x, \quad q_2\,=\,\dot x, \quad q_3\,=\,\ddot x
$$
\begin{equation}
p_1\,=\,-\,\frac{\partial L}{\partial \dot x}\,-\, \dot p_2,
\quad p_2\,=\,-\,\frac{\partial L}{\partial \ddot x}
\,-\,\dot p_3,\quad p_3\,=\,-\,\frac{\partial L}{\partial
\stackrel{\ldots}{x}}\,{,}
\end{equation}
where $L$ is the Lagrangian function in (3.4).

The action (3.4) is invariant under reparametrization $\tau \longrightarrow
f(\tau )$. Hence, the Lagrangian in (3.4) is singular or degenerated
and, as a consequence, the phase space should be restricted by constraints.

In paper [3] it has been shown  that on introducing
the canonical variables (3.5) the Hamiltonian formalism for theories
with higher derivatives is constructed completely analogous to the
Dirac generalized
Hamiltonian  dynamics dealing with singular Lagrangians depending only on
the coordinates and velocities.\footnote{In paper [5] the action (3.9) has
been cast at the beginning into an equivalent form without higher derivatives
 and then the Hamiltonian formalism has been developed}

The lagrangian in (3.4) is linear
in $\stackrel{\ldots}{x}^\mu $ therefore the definition
of the canonical momenta $p_3^{\mu }$ is a constraint itself
\begin{equation}
\stackrel{(1)}{\varphi }_\mu \,=\,p_{3\mu }\,+\,\beta
\,\frac{\sqrt{q_2^2}}{g}\,
\varepsilon_{\mu \nu \lambda }\,q_2^\nu \,q_3^\lambda \,\approx \,0\,{,}
\end{equation}
$$
\mu ,\,\nu ,\, \lambda \,=\,0, \,1,\,2\,{,}
$$
where $g=(q_2\,q_3)^2\,-\,q_2^2\,q_3^2$ and sign $\approx $ means
weak equality [6].

According to Ostrogradsky, the canonical Hamiltonian is
\begin{equation}
H\ =\ -p_1 \,\dot x\,-\,p_2\,\ddot x\,-\,p_3\stackrel {\ldots }{x}\,-\,L\,
=\,-p_1\,q_2\,-\,p_2\,q_3\,+\,
m\,\sqrt{q_2^2}\,+\,\alpha \,\frac {\sqrt {g}} { q^2_2} \,{.}
\end{equation}

The Poisson brackets are defined in a standard way
\begin{equation}
  (f,\ g)\ =\ \sum_{a=1}^{3}\  \left (\frac{\partial f}{\partial p^\mu _a}
\frac{\partial g}{\partial q_{a\mu }}\  -\  \frac{\partial
f}{\partial q^\mu _a}
\frac{\partial g}{\partial p_{a\mu }} \ \right ) \ .
\end{equation}

The evolution of the model under consideration is
determined by a total Hamiltonian
\begin{equation}
H_T\,=\,H\,+\,\sum_{\mu =0}^{2}\lambda ^\mu
\stackrel{(1)}{\varphi }_\mu\,{,}
\end{equation}
where $\lambda ^\mu ,\;\;\mu =0,\,1,\,2$ are the Lagrange multipliers.

 The primary constrains are  mutually in involution in a strong sense
\begin{equation}
\left ( \fa_\mu ,\,\fa_\nu  \right )\,=\,0
\end{equation}
 The requirement of preserving the primary constraints
under time  evolution
\begin{equation}
\frac{d\fa_\mu }{d\tau }\,=\,\left ( \fa_\mu,\,H_T \right )\,\approx
0,\quad \mu \,=\,0,\,1,\,2
\end{equation}
results in the  three secondary constraints
$$
\stackrel{(2)}{\varphi }_\mu \,=\,p_{2\mu }\,-\,\frac{\alpha
}{q_2^2\sqrt{g}}\,
[\,(q_2\,q_3)\,q_{2\mu }\,-\,q_2^2\,q_{3\mu }\,]\,+\,
$$
\begin{equation}
+\beta \,\varepsilon_{\mu \nu \lambda }\,q_2^\nu \,q_3^\lambda \,
\frac{(q_2\,q_3)}{g\,\sqrt{q_2^2}}\,\approx \,
0\,{,}\quad \mu\,=\,0,\,1,\,2\,{.}
\end{equation}
Imposing the stationarity condition on constraints (3.12) one derives
$$
\frac{d\fb_\mu }{d \tau}  =  \left ( \fb_\mu,\,H_T  \right )\,=
\,p_{1\mu }\,+\,m\,\frac{q_{2\mu }}{\sqrt{q^2_2}}\,-\,\beta\,
\frac{\varepsilon_{\mu \nu \lambda}q_2^\nu q_3^\lambda}{q^2_2\sqrt{q^2_2}}\,+
$$
\begin{equation}
+\,\sum_{\nu =0}^{2}\left ( \fb_\mu ,\,\fa_\nu \right )\,\lambda^\nu
\,\approx\,0,\quad
\mu \,=\,0,\,1,\,2.
\end{equation}

By rather involved calculations it can be shown that
\begin{equation}
\left( \fb_\mu ,\,\fa_\nu  \right )\,=\,\frac{\alpha }{\sqrt{g}}\,
b_\mu b_\nu ,
\end{equation}
where $b_\mu $ is a unit space-like vector directed along
the binormal of the world
curve
\begin{equation}
b_\mu \,=\,\frac{\varepsilon _{\mu \nu \rho }q_2^\nu q_3^\rho }{\sqrt {g}},
\quad b_\mu b^\mu \,=\,-1.
\end{equation}

Projecting (3.13) onto $q_2^\mu ,\;\;q_3^\mu $ and taking
into account (3.15) we
obtain two constraints of the third generation
\begin{eqnarray}
\fc _1&=& p_1q_2\,-\,m\sqrt{q_2^2}\, \approx \, 0, \nonumber \\
\fc _2 &=& p_1q_3 \,-\,m \frac{q_2 q_3}{\sqrt {q_2^2}} \, \approx \, 0.
\end{eqnarray}
Projection of (3.13) onto $b^\mu $ gives the relationship between
the Lagrange multipliers
\begin{equation}
p_1 b \,-\,\frac{\beta }{q_2^2}\, \left ( \frac{g}{q_2^2}  \right )^{1/2}
+\,\frac{\alpha }{\sqrt {g}}\, b \lambda\, \approx \,0.
\end{equation}
Differentiating the constraints (3.16) with respect to $\tau $ one obtains
\begin{eqnarray}
\frac{d \fc_1}{d \tau} &=& \left (\fc_1,\,H_T \right  )\,=\,\fc_2\,
\approx \,0,  \\
\frac{d \fc_2}{d \tau} &=& \left( \fc_2,\, H_T \right ) \,=\,m\,
\frac{g}{(q_2^2)^{3/2}}
\,+\,(\lambda b)\,(p_1 b)\, \approx \,0.
\end{eqnarray}

Thus we have two equations (3.17) and (3.19)  for two unknown
quantities $(p_1 b)$ and  $(\lambda b)$. The exact solutions
to these equations will not be required further because
we concentrate now upon the relation in the model under
consideration between the mass of the particle $M^2\,=\,p^2$ and its spin.
 When  (3.16) is taking into account,  the energy-momentum
vector $p_1^\mu $ assumes the form
   \begin{equation}
p_1^\mu \,=\,m\,\frac{q_2^\mu }{\sqrt{q_2^2}}\,-\,(p_1 b)\,b^\mu .
\end{equation}
This vector is conserved under the time evolution
as $(p^\mu _1,\,H_T)\, = \,0$.
On squaring (3.20), we have
\begin{equation}
M^2\,=\,p^2_1\,=\,m^2\,-\,(p_1b)^2.
\end{equation}
In case of the three-dimensional space-time the spin of the particle
is defined by
\begin{equation}
S\,=\,\frac{1}{2\sqrt{|p_1^2|}}\,\varepsilon_{\mu \nu \lambda}\,p_1^{\mu}\,
M^{\nu \lambda},
\end{equation}
where $M_{\mu \nu}$  are the Lorentz generators
\begin{equation}
M_{\mu \nu}\,=\,\sum_{a=1}^{3}(\,q_{a\mu}\,p_{a\nu }\,-
\,q_{a\nu }\,p_{a\mu}\,).
\end{equation}
When substituting (3.23) into (3.22) the spin of the particle becomes
\begin{equation}
S\,=\,\frac{1}{\sqrt{| p_1^2|}}\,\varepsilon_{\mu \nu \lambda}\,p_1^\mu \,
(\,q_2^\nu \,p_2^\lambda\,+\,q_3^\nu \,p_3^\lambda \,)\,{.}
\end{equation}
Now let us calculate $S$ on the submanifold of the phase space defined by
the constraint equations (3.6), (3.12) and by expansion (3.20).
As a result we derive
\begin{equation}
S\,=\,\pm\,\alpha \sqrt{\mu ^2\,-\,\varepsilon}\,-\,\beta \,\mu,
\end{equation}
where  $\mu \,
=\,m/\sqrt{|p_1^2|}\,\geq\,1, \quad \varepsilon\,=\,\mbox{sign} \,p_1^2$.
Thus, the Regge trajectory is split into two branches,
i.e., the mass  being fixed,  there are two states with
different spin values.
As it will be shown further, it is just  this peculiarity of the spectrum
that will be lost in dealing with squared primary constraints.

\setcounter{section}4
\setcounter{equation}0
{\bf 4.} In this section we construct the Hamiltonian formalism in
the model under consideration
starting, instead of (3.6), with squared primary constraints
\begin{eqnarray}
\ffa _1 &=& p_3^2\,+\,\beta^2\,\frac{q_2^2}{g}\,\approx \,0, \\
\ffa _2 &=& p_3q_2 \,\approx \,0, \\
\ffa _2 &=& p_3q_3 \,\approx \,0.
\end{eqnarray}
The constraint $\ffa_1$ is obtained by moving the second term in
(3.6) into the right-hand side and by squaring this equation.
The constraints (4.2)
and (4.3) are the projections of (3.6) onto $q_2$ and $q_3$, respectively.

The canonical Hamiltonian (3.7) remains, obviously, the same but the  total
Hamiltonian $\bar H_T$ is
constructed now with primary constraints (4.1) -- (4.3)
\begin{equation}
\bar H_T\,=\,H\,+\,\sum_{a=1}^{3}\mu ^a \ffa_a.
\end{equation}
Here $\mu ^a, \;\; a\,=\,1,\,2,\,3$ are new Lagrange multipliers.

The primary constraint (4.1) -- (4.3) are mutually
in involution in  a weak sense
$$
\left ( \ffa _1,\,\ffa _2 \right )\,=\,0, \quad
  \left ( \ffa _1\,\ffa _3 \right )\,=\,2\ffa _3\,\approx \,0,
$$
\begin{equation}
  \left ( \ffa _2,\,\ffa _3 \right ) \,=\,\ffa_2\,\approx \,0.
\end{equation}
The requirement of preserving the primary constraints
(4.1) -- (4.3) under time evolution results in
the three secondary constraints
\begin{eqnarray}
\ffb _1 &=& p_2p_3\,-\,\beta ^2\,\frac{q_2q_3}{g}\,\approx \,0, \\
\ffb _2 &=& p_2 q_2 \,\approx \,0, \\
\ffb _3 &=& p_2q_3\,-\,\alpha\, \frac{\sqrt{g}}{q_2^2}.
\end{eqnarray}

The constraints (4.6) -- (4.8) are in a complete
agreement with constraints (3.12).
Really, projecting (3.12) onto (3.6), $q_2^\mu $ and $q_3 ^\mu $
we arrive at the
constraints (4.6) -- (4.8).

Disagreement appears in the following. The constraints (4.6) --
(4.8) turn out to be
in involution in a
weak sense with primary constraints (4.1) -- (4.3)
\begin{equation}
 \left (  \ffb_a,\,\ffb_b \right ) \,\approx \,0, \quad
a,\,b \,=\,1,\,2,\,3,
\end{equation}
while the constraints (3.12) and (3.6) do not  commute (see eq.\ (3.14)).

Differentiating constraints (4.6) -- (4.8) with respect to $\tau $,
\begin{equation}
\frac{d \ffb_a}{d\tau }\,=\,\left ( \ffb_a,\,\bar H_T \right )\,\approx \,
\left ( \ffb_a,\, H \right )\,\approx \,0, \quad a\,=\,1,\,2,\,3
\end{equation}
three new constraints are derived
\begin{eqnarray}
\ffc _1 &=& p_1p_3\,+\, p_2^2\,+\,\alpha\,\frac{p_2q_3}{\sqrt{g}}\,
+\,\beta ^2\,\frac{q_3^2}{g}
\,\approx \,0, \\
\ffc _2 &=& p_1q_2\,-\,m\,\sqrt{q_2^2}\,\approx \,0,\\
\ffc _3 &=& p_1q_3\,-\,m\,\frac{q_2q_3}{\sqrt{q_2^2}}\,\approx \,0.
\end{eqnarray}
Constraints (4.12) and (4.13) are completely equivalent to (3.16) but
the constraint (4.11) has no counterpart between the
constraints derived in the preceding section.

The requirement of the stationarity of the constraints (4.11) --
(4.13) enables one
to fix two Lagrangian multipliers $\mu _1$ and $\mu _3$ while
$\mu _2$ remains arbitrary [4].

 It is convenient to use further the proper time gauge
\begin{equation}
q_2^2 \,=\,\mbox{const}, \quad q_2q_3\,=\,0.
\end{equation}
In this case, three vectors $q_2,\;q_3,$ and $p_3$ form, owing to
(4.1) -- (4.3) and
(4.14) a complete orthogonal basis. From () 4.6 -- (4.8) we deduce
\begin{equation}
p_2^\mu \,=\,-\alpha \,\frac{q_3^\mu }{\sqrt{-q_2^2q_3^2}}.
\end{equation}
Given (4.11) -- (4.13), we can derive in the same way
\begin{equation}
p_1^\mu \,=\,m\,\frac{q_2^\mu}{\sqrt{}q_2^2}\,+\,p_3^\mu
\,\frac{q_3^2}{q_2^2}.
\end{equation}
For the mass squared this yields
\begin{equation}
M^2\,=\,p_1^2\,=\,m^2\,+\,\beta ^2\,\frac{q_3^2}{(q_2^2)^2}\,
=\,m^2\,-\,\beta ^2\,k^2(s)
\end{equation}
instead of (3.21)
This expression, as well as (3.21), is not positive definite because of
$q_3^2 \,<\,0$ and $k^2(s)\,>\,0$.

Let us calculate the spin  of the particle according to
(3.24). We should  here evaluate a quantity $V\,=\,\varepsilon
_{\mu \nu \lambda}
q_2^\mu q_3^\nu p_3^\lambda $ on the physical submanifold of the
phase space. By making use of the primary constraints in the form (4.1) --
(4.3) we
can find $V$ up to  sign and, as a consequence,
the spin of the particle will be determined up to sign. To remove
this ambiguity
we fix the sign
of $V$ using the calculations of the preceding section, which  gives
\begin{equation}
V\,=\,-\beta \,\sqrt{q_2^2}.
\end{equation}
 Finally, the particle spin is given by
\begin{equation}
S\,=\,\alpha \,\sqrt{\mu ^2\,-\,\varepsilon}\,-\,\beta \,\mu ,
\end{equation}
where $\mu $ and $\varepsilon $ are the same parameters as in eq.\ (3.25).

Thus, dealing with squared primary constraints
(4.1) -- (4.3) we have lost the two-valuedness of the Regge trajectory.

\setcounter{section}5
\setcounter{equation}0
{\bf 5.} In conclusion it should be noted the following. The squared
primary constraints
(4.1) -- (4.3) appear inevitably when treating the action (3.1) in the
space-time with
dimension $D >3$. In this case the torsion of the world curve is determined
not by eq.\ (3.3), linear in $\stackrel{\ldots }{x}$, but by the nonlinear
expression
\begin{equation}
\kappa  (s) \,=\,\frac{\sqrt{\det (d_{\alpha \beta})}}{k^2(s)}\,{,}
\end{equation}
$$
d_{\alpha \beta }\,=\,\stackrel{(\alpha )(\beta )}{x^\mu x_\mu},
\quad \stackrel{(\alpha )}{x}\,\equiv \,d^\alpha x/ds^\alpha ,\quad
\alpha ,\,\beta \,=\,1,\,2,\,3.
$$
The definition (5.1) makes  sense for $D=3$ too. In this case it
gives an absolute value of the torsion defined
by (3.3). The action (3.1) with torsion given by (5.1) has  been
considered in [4]
and the mass spectrum (4.19) squared has been derived there.

Thus, the use of primary constraints  in the squared form (4.1) -- (4.3)
results really  in replacing  the model (3.1), (3.3) by (3.1), (5.1).
It has been shown recently in non-manifest way in paper [16] where the
model (3.1), (3.3)
was treated by making use of the squared primary constraints (4.1)--(4.3)
in the total Hamiltonian.

\vskip0.5cm
{\large \bf Acknowledgement}
\vskip0.3cm
The author is grateful to M.\ S.\ Plyschchay for valuable
discussions of problems
treated in this paper and for parallel calculation of the Poisson
brackets in section 3.

\vfill
\begin{center}
Received by Publishing Department\\
on September 2,  1993
\end{center}

\end{document}